\documentclass[floatfix,twocolumn,aps,prd,showpacs,amsmath,amssymb]{revtex4}
\usepackage{epsfig}
\usepackage[latin1]{inputenc}

\begin{document}

\title{Chiral-Symmetry Breaking  in Pseudo Quantum Electrodynamics at Finite Temperature}

\author{Leandro O. Nascimento$^{1,2,3}$, Van S\'ergio Alves$^{1,2}$, Francisco Pe\~na$^4$, C. Morais Smith$^3$, and E. C. Marino$^2$}
\affiliation{$^1$ Faculdade de F\'\i sica, Universidade Federal do Par\'a, Av.~Augusto Correa 01, 66075-110, Bel\'em, Par\'a, Brazil \\
$^2$ Instituto de F\'\i sica, Universidade Federal do Rio de Janeiro, C.P.68528, Rio de Janeiro RJ, 21941-972, Brazil \\
$^3$Institute for Theoretical Physics, Center for Extreme Matter and Emergent Phenomena Utrecht University, Leuvenlaan 4, 3584CE Utrecht, The Netherlands \\
$^4$ Departamento de Ciencias F\'\i sicas, Facultad de Ingenier\'\i a, Ciencias y Administrac\'\i on, Universidad de La Frontera, Avda. Francisco Salazar 01145, Casilla 54-D, Temuco, Chile}

\date{\today}

\begin{abstract}
We use the Schwinger-Dyson equations in the presence of a thermal bath, in order to study chiral symmetry breaking in a system of massless Dirac fermions interacting through pseudo quantum electrodynamics (PQED3), in (2+1) dimensions. We show that there is a critical temperature $T_c$, below which chiral symmetry is broken, and a corresponding mass gap is dynamically generated, provided the coupling is above a certain, temperature dependent,
critical value $\alpha_c$.~The ratio between the energy gap and the critical temperature for this model is estimated to be $2 \pi$.~These results are confirmed by analytical and numerical investigations of the Schwinger-Dyson equation for the electron. In addition, we calculate the first finite-temperature corrections to the static Coulomb interaction. The relevance of this result in the realm of condensed matter systems, like graphene, is briefly discussed. 
\end{abstract}


\maketitle

\section{Introduction}
Chirality is a physical quantity, carried by fermionic particles, which characterizes how these behave under a specular reflection. Chiral symmetry is the invariance under an operation relating different chiralities. It may be either discrete or continuous but, in any case, masslessness is usually a necessary condition for a system to be chiral symmetry invariant. Chiral symmetry is usually broken at a quantum level by radiative corrections, which are non-perturbative in the coupling constant.

A very convenient platform to investigate chiral symmetry breaking is the non-perturbative approach provided  by the Schwinger-Dyson equations (SDE). This approach consists of a complicated system of integral equations relating exact Green's functions and vertex functions \cite{Roberts}. Nontrivial solutions of the SDE usually imply a dynamical mass generation for the matter field. The case of quantum electrodynamics in (3+1) dimensions (QED4) has been studied in Ref.~\cite{Maris3}. For quantum electrodynamics in (2+1) dimensions (QED3), the chiral-symmetry breaking for massless Dirac fermions has been extensively studied, both at zero \cite{Roberts, Appel, Kondo, Maris2, Hong} and finite temperature \cite {Dorey, Ayala, Aitchison,Triant}.
  
Three-dimensional theories, such as QED3, have recently attracted much attention because these models are relevant for the theoretical description of some effects typically observed in condensed-matter systems. Examples range from high-T$_c$ superconductivity \cite{Roberts,Oskar} to graphene \cite{GeimIQHE,MariaVoz,Drut,ChunXu,VLWJF}.~Graphene is particularly interesting because its linear tight-binding dispersion relation
coincides with the one of massless Dirac particles. Furthermore, 
the system has particle-hole symmetry \cite{review}.    

However, even in two-dimensional materials the real photons propagate in (3+1) dimensions. This raises the question of how to describe the electromagnetic interactions of charged particles in (2+1) dimensions. This problem was studied in Refs.~\cite{DoreyNL,Kovner,marino}. After projecting the photons onto (2+1) dimensions, a non-local term emerges in the Maxwell Lagrangian, leading to the so-called pseudo quantum electrodynamics (PQED) \cite{marino,marino1}, sometimes also referred to as reduced quantum electrodynamics \cite{Teber,Miransky0}. 

It has been shown that PQED correctly describes the physical 1/r Coulomb interaction between static charges, contrarily to QED3 at tree level, which provides a logarithmic static interaction $\ln(e^2 r)$.~The processes of canonical quantization of theories such as PQED may be found in the Refs~\cite{Barci,Marino2,Barcelos}. There, it was shown that PQED respects causality. Furthermore, it has been shown that PQED respects the Huygens principle in (2+1) dimensions, unlike QED3 \cite{huygens}.~More recently, it has been also proved that this theory respects unitarity \cite{unitarity}.

In a previous publication, we investigated the occurrence of chiral-symmetry breaking in PQED at T=0 \cite{VLWJF}.
In this paper, we study the effect of temperature on the dynamical chiral-symmetry breaking
in PQED coupled to massless Dirac fermions, by considering this theory in the presence of a thermal bath. 
At the classical level, the theory is chiral invariant due to the absence of a mass term for the fermions.
We show that there exists a critical temperature $T_c$, below which quantum effects produce dynamical
breakdown of the chiral symmetry, with the associated generation of a mass gap.
This occurs above a critical coupling $\alpha_c(T)$, which is temperature dependent and increases with temperature. The temperature, therefore, is an inhibitor of dynamical mass generation. We estimate the ratio between the mass gap and $T_c$ to be approximately $2 \pi$. All these results are then confirmed by numerical calculations.
Finally, we evaluate the corrections to the static Coulomb potential of PQED, due to the finite temperature.
For this purpose, we use the SDE for the gauge-field propagator, which involves the vacuum polarization. We find a logarithmic correction in the limit of short-range distance. For the long-range limit, the corrected interaction is proportional to the inverse of the third power of the distance.
    
The outline of this paper is the following: In Sec.~II we review the Feynman rules for PQED, while in Sec.~III we study the electron self-energy in PQED by using the SDE and the imaginary-time formalism or Matsubara frequencies. In Sec.~III.A we prove that the sum over Matsubara frequencies is convergent for PQED, thereafter we solve this sum;
in Sec.~III.B we use the zero-mode approximation in order to solve the SDE analytically, and in Sec.~III.C we use the zero-external-momentum approach in the mass function. We obtain the critical temperature, the critical coupling constant, and we also estimate the ratio between the energy gap and the critical temperature. Then, we discuss how to include more Matsubara frequencies in the calculations. In Sec.~IV we calculate the corrections to the Coulomb potential in (2+1) dimensions due to the thermal bath. We also include two appendices, in the  first one we compare our analytical results to the  numerical solution of the full integral equation for the mass function, while in the second one, we discuss the approach about the wave function renormalization.

\section{Chiral-Symmetry breaking in PQED at $T=0$}

The Lagrangian of the PQED in Euclidean space-time and for massless Dirac fermions is given by \cite{marino}
\begin{eqnarray}
{\cal L}=\frac{1}{4}F^{\mu\nu}\frac{2}{(-\Box)^{1/2}}F_{\mu\nu}+
\bar\psi(i\gamma^\mu\partial_\mu+e\,\gamma^\mu A_\mu)\psi, \label{PQED3}
\end{eqnarray}
where $\Box$ is the d'Alembertian operator, $e$ is the electric charge of the electron, $F_{\mu\nu}$ is the usual field intensity tensor of the U(1) gauge field $A_\mu$, the pseudo electromagnetic field, $\psi$ is a four-component Dirac field, and $\gamma^{\mu}$ are rank-4 Dirac matrices. 

The Feynman rules, at zero temperature, imply that the gauge-field propagator reads \cite{VLWJF} 
\begin{equation}
\Delta_{0\mu\nu}(p)=\frac{1}{2\sqrt{p^2}}\left(\delta_{\mu\nu}-\frac{p_\mu p_\nu}{p^2}\right) \label{propA}
\end{equation}
in the Landau gauge and that the inverse of the fermion propagator is
\begin{equation}
S^{-1}_{0F}(p)=-\gamma^\mu p_\mu \label{proppsi}.
\end{equation} 

Recently, the authors in Ref.~\cite{VLWJF} investigated chiral symmetry breaking in PQED at zero temperature using the quenched- and the unquenched-rainbow approaches to solve the SDE ( see Ref.~\cite{Roberts} for a insightful review about the SDE). In the quenched-rainbow approach, the vacuum polarization is neglected, whereas in the unquenched one it is considered. Within the first approach, chiral-symmetry breaking only occurs if $\alpha>\alpha_c$, whereas for the second, a value for the critical number of fermions assets given by $ N_c(\alpha)$ was obtained. In this case, chiral symmetry breaking only occurs if $N_f<N_c(\alpha)$. Furthermore, it has been shown that the model has intrinsic characteristics of both, QED4 (renormalizable theory) and QED3. Specially, for QED3 the critical number of active fermions is reobtained from PQED in the strong coupling limit, i.e, $N_c(\alpha\rightarrow\infty)=N_c^{QED3}=32/\pi^2$. For QED3 this value was first obtained in Ref.~\cite{Appel}.

\section{Quenched-rainbow approximation at finite temperatures}\label{sec3}

The chiral-symmetry breaking in QED4 at finite temperatures was obtained from non-trivial analytical solutions for the mass function $\Sigma(p)$, which only appear below a certain cri\-ti\-cal temperature $T_c$, estimated using the zero mode of the Matsubara frequencies \cite{Ayala}. Above the critical temperature, the chiral symmetry is restored. It has been also shown that the integral equation for the mass function in QED4 accounting for temperature effects is the same as in QED3 by taking $T=0$ and using a dimensional coupling constant $\gamma=e^2T/(4\pi)=\alpha T$, fixed at the limit $T\rightarrow 0$. In addition, this analytical approach for QED4 has been also confirmed by a numerical solution of the full Schwinger-Dyson equation for the mass function \cite{Ayala}. However, the same set of approximations is not applicable to QED3 because there are infrared divergences in the electron self-energy if the vacuum polarization is not taken into account.

Due to the similarities between PQED and QED4, here we use the same set of approaches proposed in Ref.~\cite{Ayala}, and as expected, no infrared divergence is observed in the SDE. The inverse of the full electron propagator then reads
\begin{equation}
S_F^{-1}(p)=S_{0F}^{-1}(p)-\Xi(p),  \label{fullpropfer2}
\end{equation}
where the electron self-energy $\Xi(p)$ is given by
\begin{eqnarray}\label{fermion}
\Xi (p)=e^2\int\frac{d^3k}{(2\pi)^3} \gamma^{\mu}S_F(k)\Gamma^{\nu}(k,p)\,\Delta_{\mu\nu}(p-k),
\end{eqnarray}
with $\Gamma^\nu(k,p)$ denoting the full vertex function. In the lowest order, we consider $\Gamma^\nu(k,p)=\gamma^\nu$, which is called rainbow approach. Furthermore, we use only the free gauge-field propagator. This assumption is called quenched approach. Both approximations are motivated because of perturbation theory, which establishes $\Gamma^\nu(k,p)=\gamma^\nu+O(e^3)$ and $\Delta_{\mu\nu}=\Delta_{0\mu\nu}+O(e^2)$ \cite{Roberts,Maris3,Appel,Kondo}.

The inverse of the full fermion propagator can be written as \cite{Roberts,Maris3,Appel,Kondo}
\begin{equation}
S_F^{-1}(p)=-p_{\mu}\gamma^{\mu}\,A(p)+\Sigma (p), \label{full1}
\end{equation}
where $A(p)$ is the identity-proportional term (this term is also called the renormalization function), and $\Sigma(p)$ is called the mass function. To obtain $\Sigma(p)$ we use Eq.~(\ref{full1}) in Eq.~(\ref{fullpropfer2}) and calculate the trace over the Dirac matrices. The corresponding result for zero temperature is \cite{VLWJF} 
\begin{equation}
\Sigma(p)=4\pi\alpha\int\frac{d^3k}{(2\pi)^3}\frac{\Sigma(k)}{A^2(k)k^2+\Sigma(k)^2}\frac{1}{|p-k|}, \label{ese1}
\end{equation}
with $\alpha=e^2/{4\pi}$ denoting the coupling constant. To obtain $A(p)$, we multiply Eq.~(\ref{fullpropfer2}) by the Dirac matrices and calculate the trace over them. The result for zero temperature is \cite{VLWJF}
\begin{eqnarray}
A(p)=1&+&\frac{4 \pi \alpha}{ p^2}\int\frac{d^3k}{(2\pi)^3}\frac{A(k)\Delta(q)}{k^2A^2(k)+\Sigma^2(k)} \nonumber
\\
&\times& \frac{(p.q)(k.q)}{q^2}, \label{waveAT}
\end{eqnarray}
where
\begin{eqnarray}
q=p-k 
\end{eqnarray}
and
\begin{eqnarray}
\Delta(q)=\frac{1}{2(q^2)^{1/2}}.
\end{eqnarray}

It has been shown that $A(p) \approx 1$ for zero temperature. Therefore, if $A(p)=1$, $\Sigma(p)$ fully describes the electron self-energy $\Xi(p)$ \cite{VLWJF}. In appendix B, we show that this result also holds for finite temperatures. Therefore, from now on we consider $A(T,\textbf{p})=1$.

Eq.~(\ref{ese1}) is very similar to Eq.~(14) in Ref.~\cite{VLWJF}. However, there is a different coefficient in the mass function due to a constant factor in the gauge-field propagator. Although this will change the critical behavior of the theory, one verifies that the correct factor in Eq.~(\ref{propA}) should be $1/2$ in order to obtain the precise Coulomb interaction between charges in the plane. Indeed, the critical coupling constant for the onset of chiral-symmetry breaking in Eq.~(\ref{ese1}) is $\alpha_c=\pi/8\approx 0.40$ for $T=0$. We rewrite Eq.~(\ref{ese1}) as two integrals, one for the energy component and another for the two-dimensional momentum components,
\begin{eqnarray}
\Sigma(p)=4\pi\alpha\int\frac{dk_0}{(2\pi)}\int\frac{d^2\textbf{k}}{(2\pi)^2}\frac{\Sigma(k)}{k_0^2+\textbf{k}^2+\Sigma(k)^2}\nonumber\\
\times \frac{1}{[(p_0-k_0)^2+(\textbf{p}-\textbf{k})^2]^{1/2}}.
\end{eqnarray}

Next, we introduce the effects of temperature through the Matsubara frequencies in the temporal component. For this purpose, we must replace $k_\mu =(k_0,\textbf{k})\rightarrow(\omega_n,\textbf{k}) $ and $p_\mu =(p_0,\textbf{p})\rightarrow(\omega_m,\textbf{p})$, where $\omega_n$ and $\omega_m $ are the Matsubara frequencies for fermions, given by $\omega_n =(2n +1)\pi T $ and $\omega_m= (2m+1)\pi T $, with $n$ and $m$ integers \cite{Abrikosov}. The integral over $k_0$ becomes a sum in $n$ by replacing $\int dk_0 f(k_0,\textbf{k},T) \rightarrow 2\pi T \sum_n f(\omega_n,\textbf{k},T)$ and the mass function is rewritten as
\begin{eqnarray}
\Sigma_m(p)&=&\sum_{n=-\infty}^{+\infty}\int\frac{d^2\textbf{k}}{(2\pi)^2}\frac{(4\pi\alpha T)\Sigma(k)}{(2n+1)^2\pi^2T^2+\textbf{k}^2+\Sigma(k)^2}\nonumber\\
&\times&\frac{1}{[4(n-m)^2\pi^2T^2+(\textbf{p}-\textbf{k})^2]^{1/2}}.  \label{sigmaT}
\end{eqnarray}

This is the integral equation for the mass function in PQED3 at finite temperatures when considering the sum over all Matsubara frequencies in the quenched-rainbow approach. We have an additional complication with respect to the zero temperature case, which is the dependence of the $\Sigma_m(p)$ function on the Matsubara frequencies $\omega_m$. We will assume that the mass function is independent of $\omega_m$, and therefore consider $m=0$. This also implies that $\Sigma(k)$ does not participate on the sum over $n$.

\subsection{The Sum Over Matsubara Frequencies in PQED}

In QED4, the sum over all Matsubara frequencies present in the integral equation for the mass function $\Sigma(p,T)$ is calculated by standard methods \cite{Ayala}. However, for PQED it is not simple to obtain the solution of the complete series. First, we need to analyze its convergence. Starting from Eq.~(\ref{sigmaT}), we see that the sum over $n$ can be expressed as a sum $U$, given by
\begin{equation}
U=\sum_{n=-\infty}^{+\infty} u_n=\sum_{n=-\infty}^{+\infty}\frac{1}{\sqrt{n^2+A^2}}\frac{1}{(2n+1)^2+B^2}.\label{U}
\end{equation}
Note that the terms of the sum are all positive, since $A^2,B^2>0$. Furthermore, we can write $U= 2\sum_{n=1}^{+\infty}u_n^{(+)}+u_0$, where $u_n^{(+)}=(u_n+u_{-n})/2$. Therefore, it is clear that if $\sum_{n=1}^{+\infty}u_n^{(+)}$ is a convergent sum, then $U$ will be too. 

Next, we define another independent sum $V$, which is
\begin{equation}
V=\sum_{n=1}^{+\infty}v_n=\sum_{n=1}^{+\infty}\frac{1}{(n+A)}\frac{1}{(n+B)}, \label{V}
\end{equation} 
where $V$ is convergent and has a known result. Furthermore, we have that $v_n$ and $u^{(+)}_n$ satisfy the inequalities $v_n\geq u^{(+)}_n\geq 0$ and
\begin{equation}
\lim_{n\rightarrow\infty}\frac{u^{(+)}_n}{v_n}=0.
\end{equation}

Therefore, we can use the comparison test of series \cite{Arfken}, which establishes that if $V$ is convergent, then $\sum_{n=1}^{+\infty}u^{(+)}_n$ is necessarily convergent, which implies that $U$ is also convergent, completing our proof. Furthermore, we have $u_0>u_n$, for any $n$.

From now on, we will not write time components anymore, hence it is needless to use bold in the momentum components. This allows us to simplify the notation, $\Sigma_0(\textbf{p},T)\equiv \Sigma(p)$.

After showing that $U$ is convergent, we follow an analytical procedure to calculate the sum over the Matsubara frequencies in Eq.~(\ref{sigmaT}). This is useful in order to identify the temperature-independent term of Eq.~(\ref{sigmaT}), which is called the vacuum term \cite{vterm}. Thereby, we rewrite Eq.~(\ref{sigmaT}) for $m=0$
\begin{equation}
\Sigma(p)=4\pi\alpha \int\frac{d^2\textbf{k}}{(2\pi)^2}  \Sigma(k)\int_{-\infty}^{+\infty} \frac{dy}{\pi} \sigma_y(p,\Sigma(k)), \label{sigma0p}
\end{equation}
where the kernel $\sigma_y$ is given by
\begin{equation}
\sigma_y=\sum_{n=-\infty}^{+\infty} \frac{T}{4n^2\pi^2 T^2+\omega_{p,k}^2}\frac{1}{(2n+1)^2\pi^2 T^2+\epsilon_{k}^2}, \label{kernel1}
\end{equation}
with
\begin{equation}
\omega_{p,k}^2=y^2+(\textbf{p}-\textbf{k})^2,
\end{equation}
and
\begin{equation}
\epsilon_k^2=k^2+\Sigma^2(k).
\end{equation}

Next, we calculate the sum over the Matsubara frequencies in Eq.~(\ref{kernel1}) and find
\begin{equation}
\sigma_y=C_B [2 n_B(\omega_{p,k})+1]-C_F [1-2n_F(\epsilon_k)] \label{matsum},
\end{equation}
where $n_B(\omega_{p,k})=[\exp(\omega_{p,k}/T)-1]^{-1}$ and $n_F(\epsilon_{k})=[\exp(\epsilon_{k}/T)+1]^{-1}$ are the Bose and Fermi distribution functions, respectively. The functions $C_B$ and $C_F$ are
\begin{equation}
C_B(T)=\frac{(s_{p,k} d_{p,k}+\pi^2T^2)}{2\omega_{p,k}(d^2_{p,k}+\pi^2T^2)(s^2_{p,k}+\pi^2T^2)},
\end{equation}
and
\begin{equation}
C_F(T)=\frac{(s_{p,k} d_{p,k}-\pi^2T^2)}{2\epsilon_k(d^2_{p,k}+\pi^2T^2)(s^2_{p,k}+\pi^2T^2)},
\end{equation}
where $d_{p,k}\equiv \epsilon_k-\omega_{p,k}$ and $s_{p,k}\equiv \epsilon_k+\omega_{p,k}$. 

We obtain the vacuum term by using $T=0$ in Eq.~(\ref{matsum}). Thereby, the kernel at zero temperature is
\begin{equation}
\sigma_y(T=0)=C_B(0)-C_F(0)=\frac{1}{2 \omega_{p,k}\epsilon_k(\omega_{p,k}+\epsilon_k)}. \label{vterm}
\end{equation}
Integrating out $y$ in Eq.~(\ref{vterm}) yields 
\begin{equation}
\frac{\pi}{2 \epsilon_k \sqrt{\textbf{q}^2-\epsilon_k^2}}\left[1-\frac{2}{\pi} \tan^{-1}\left(\frac{\epsilon_k}{\sqrt{\textbf{q}^2-\epsilon^2_k}}\right)\right],
\end{equation} 
where $\textbf{q}=\textbf{p}-\textbf{k}$. Therefore, Eq.~(\ref{sigmaT}) becomes
\begin{equation}
\Sigma(p)|_{T=0}=4 \pi \alpha  \int \frac{d^2 \textbf{k}}{(2\pi)^2}\frac{\Sigma(k)}{k^2+\Sigma^2(k)}G(\textbf{q},\Sigma(k)), \label{finmat}
\end{equation}
where
\begin{equation}
G=\frac{\epsilon_k}{2\sqrt{\textbf{q}^2-\epsilon^2_k}}\left[1-\frac{2}{\pi} \tan^{-1}\left(\frac{\epsilon_k}{\sqrt{\textbf{q}^2-\epsilon^2_k}}\right)\right].
\end{equation}

Admitting $\Sigma(k)$ is finite for $k\rightarrow \infty$, it follows that Eq.~(\ref{finmat}) is logarithmically divergent. This is the same degree of divergence of the mass-proportional term in the electron self-energy, obtained through perturbation theory at one-loop expansion at zero temperature. Here, however, since we are interested in the non-perturbative regime, we shall consider $\Sigma(k)$ a non-trivial function of the momentum, which vanishes at large momentum. For this purpose, it is more convenient to convert the integral equation into a differential equation with asymptotic conditions. 

Although the previous analytical method is sufficient to solve the sum over all Matsubara frequencies, the angular integration obtained from Eq.~(\ref{sigma0p}) and Eq.~(\ref{matsum}) does not allow us to obtain the differential equation for the mass function. Next, we shall explore a different analytical approach to circumvent this problem.

We return to the integral equation, but we integrate out the angular variable instead of summing over the Matsubara frequencies. Therefore, Eq.~(\ref{sigmaT}) reads
\begin{equation}
\Sigma(p)=\frac{16\pi\alpha T}{(2\pi)^2}\int_0^{\infty}kdk\Sigma(k)I(p,k,\Sigma(k)),\label{sigmaTn}
\end{equation}
where $I$ is given by
\begin{eqnarray}
I(k,p,\Sigma(k))=\sum_{n=-\infty}^{+\infty}\frac{1}{(2n+1)^2\pi^2T^2+k^2+\Sigma(k)^2}\nonumber\\
\times\frac{K(x_n)}{\sqrt{(p-k)^2+4n^2\pi^2T^2}}, \label{I}
\end{eqnarray}
with $K(x_n)$ denoting the elliptic function. For $x_n<1$, we have $K(x_n)=\pi/2+\pi x_n/8+O(x_n^3)$, with $x_n$ given by
\begin{equation}
x_n=-\frac{4pk}{(p-k)^2+4n^2\pi^2T^2}.
\end{equation}

Next, we propose an approximation to the sum $U$, in order to obtain an analytical solution to $\Sigma(p)$. The zero-mode approximation, i.e,  $U\approx u_0$ has been used for QED4 in Ref.~\cite{Ayala} and is also valid for the PQED.

\subsection{Zero-Mode Approximation}

For low temperatures, the sum over $n$ in Eq.~(\ref{I}) is approximately given by the term $u_0$. Therefore, by taking $n=0$ in Eq.~(\ref{sigmaTn}), we find
\begin{equation}
\Sigma(p)=\frac{16\pi \gamma}{(2\pi)^2}\int_0^{\infty}dk\frac{k}{|p-k|}\frac{K(x_0)\Sigma(k)}{\pi^2 T^2+k^2+\Sigma(k)^2}, \label{sigmaT2}
\end{equation}
where $\gamma=\alpha T$ is a dimensional coupling constant. This condition is necessary to study the dynamical mass generation in the case of a nonzero critical temperature $T_c$. However, one may always return to the coupling constant and consider it in the context of a finite critical coupling constant $\alpha_c(T)$ at finite temperatures. The symmetry is fully restored if $T_c=0$ or $\alpha_c(T)\rightarrow\infty$. Note that both interpretations are equivalent, in the sense that at very low temperatures, interaction effects become more relevant.

Next, we divide Eq.~(\ref{sigmaT2}) into two parts, the infrared part $k<p$ and the ultraviolet part $k>p$. Furthermore, we include a ultraviolet ``cutoff'' $\Lambda$ (bear in mind that we may return to the continuum limit $ \Lambda = \infty $ at any time in the calculations). However, for comparison with lattice field theory, it is interesting to keep the cutoff finite. Using these approaches in the elliptic function, we obtain
\begin{equation}
\frac{k}{|p-k|}K(x_0)\approx \frac{k\pi}{2p}\theta(p-k)+\frac{\pi}{2}\theta(k-p) \label{kernelPQED3}. 
\end{equation}

Therefore, Eq.(\ref{sigmaT2}) becomes
\begin{eqnarray}
\Sigma(p)=\frac{2\gamma}{p}\int_0^pdk \frac{k\Sigma(k)}{\pi^2T^2+k^2+\Sigma(k)^2}\nonumber\\
+2\gamma\int_p^\Lambda dk \frac{\Sigma(k)}{\pi^2 T^2+k^2+\Sigma(k)^2}. \label{tiv}
\end{eqnarray}

We can neglect the nonlinear term and approximate $\Sigma^2(k)+k^2+\pi^2T^2\approx k^2+\pi^2T^2$. As argued earlier in the literature, this approximation does not change significantly the critical behavior of the theory \cite{Maris2}. Furthermore, $T$ is dominant in the region where $k\rightarrow 0 $ and is negligible in the region where $ k \rightarrow \Lambda $. This shows that the temperature behaves like a natural infrared cutoff for Eq.~(\ref{tiv}). This infrared cutoff is identified as the critical temperature for chiral symmetry breaking \cite{Ayala}. Thereby, by taking the first derivative of the above expression, we have
\begin{equation}
\frac{d\Sigma(p)}{dp}=-\frac{2\gamma}{p^2}\int_{\pi T_c}^pdk \frac{\Sigma(k)}{k},
\end{equation} 
which can be converted to a second-order differential equation given by
\begin{equation}
\frac{d}{dp}\left(p^2\frac{d\Sigma(p)}{dp}\right)+\frac{2\gamma}{p}\Sigma(p)=0, \label{eqdif}
\end{equation}
with infrared and ultraviolet boundary conditions given, respectively, by
\begin{equation}
\lim_{p\rightarrow\pi T_c} p^2\frac{d\Sigma(p)}{dp}=0  \label{IRC}
\end{equation}
and
\begin{equation}
\lim_{p\rightarrow \Lambda}\Sigma(p)=0. \label{UVC}
\end{equation}

The solution of the differential Eq.~(\ref{eqdif}) is
\begin{eqnarray}
\Sigma(p)=C_1 \sqrt{\frac{\gamma}{p}}J_1\left(\sqrt{\frac{8\gamma}{p}}\right)
+C_2 \sqrt{\frac{\gamma}{p}}Y_1\left(\sqrt{\frac{8\gamma}{p}}\right), \label{soleqdif}
\end{eqnarray}
where $J_1(x)$ and $Y_1(x)$ are the Bessel functions of first and second kind, respectively. $C_1$ and $C_2$ are arbitrary constants with dimension $[C_1]=[C_2]=1$, both in units of energy. From Eq.~(\ref{UVC}), it is possible to show that $C_2=0$, because $x \,Y_1(x)$ is not zero if $x\rightarrow 0$, with $x=\sqrt{\gamma/\Lambda}$. On the other hand, Eq.~(\ref{IRC}) provides the identity
\begin{equation}
2J_1(\xi)+\xi J_0(\xi)-\xi J_2(\xi)=0, \label{IVPQED3}
\end{equation}
where $\xi = \sqrt{8\gamma /{\pi T_c}}$. Eq.~(\ref{IVPQED3}) has a set of nontrivial solutions $\{\xi_n\}=\{\xi_0,\xi_1,...\}$, where $\xi_n<\xi_{n+1}$ for any $n$. The minimal value $\xi_0$ is chosen to define the critical temperature $T_c$,
\begin{equation}
T_c=\frac{8 \gamma}{\pi \xi_0^2}. \label{TcdaPQED}
\end{equation}

Using Eq.~(\ref{IVPQED3}), one can show that $\xi_0 \approx 2.40$, thus allowing us to determine the critical temperature only in terms of the dimensional coupling constant, $T_c =8 \alpha T/\pi \xi_0^2=0.44 \,\alpha T$, i.e, for $T<T_c$, then $\alpha>\alpha_c(T \approx T_c)= \pi \xi_0^2/8 \approx 2.26$. This result shows that even for low temperatures the critical coupling constant is much larger than in the case of zero temperature $\alpha_c(T=0)=\pi/8\approx 0.40$ \cite{VLWJF}. 

For an arbitrary value of $\gamma$, it can be verified by numerical tests that the full integral equation, given by Eq.~(\ref{sigmaT2}), obeys $\Sigma(p,T\geq T_c)\rightarrow 0$. Therefore, a nontrivial solution for $\Sigma(p)$ requires a nontrivial value for $T_c$. 

The approach of this section has provided an analytical solution for the mass function $\Sigma(p,T)$, as well as an estimative for $T_c$. Unfortunately, from the analytical solution, it is not clear that $\Sigma(p,T=T_c)\rightarrow 0$. In order to confirm the phase transition at $T_c$ and $\alpha_c(T)$, in the next section, we will recalculate the mass function, but at zero external momentum. 

\subsection{Zero-External-Momentum Approximation}

In this section, we will take zero external momentum in Eq.~(\ref{sigmaT2}), i.e, $\Sigma (p,T) \approx \Sigma (0,T)= m(T)$. At zero momentum, the mass function is the pole of the fermion propagator, therefore $2\,m(T)$ is the energy gap between positive and negative energies. It has been argued that the mass function reaches its maximum value exactly at $p\rightarrow 0$ \cite{Maris2,VLWJF}. Therefore, if $m(T)\neq 0$ we immediately conclude that chiral-symmetry breaking will occur. By using this approach into Eq.~(\ref{sigmaT2}), we have
\begin{equation}
1=2\gamma\int_0^\Lambda \frac{dk}{\pi^2T^2+k^2+m(T)^2}. \label{eqm}
\end{equation}

Eq.~(\ref{eqm}) can be transformed into a transcendental equation for $m(T)$, given by
\begin{equation}
\frac{\sqrt{m(T)^2+\pi^2 T^2}}{2\gamma}=\tan^{-1}\left(\frac{\Lambda}{\sqrt{m(T)^2+\pi^2T^2}}\right).  \label{mTn0}
\end{equation}

In order to obtain an analytical solution for $m(T)$, we expand the right-hand side (rhs) of Eq.~(\ref{mTn0}) for $\Lambda > \sqrt{m(T)^2+\pi^2T^2}$. This is a reasonable assumption, since we are in the low-temperature regime and the generated mass is expected to be much smaller than the ultraviolet cutoff. We perform a Taylor expansion until second order, $\tan^{-1}(x)\approx \pi/2-1/x+O(x^{-2})$ if $x>1$, and we find two possible solutions
\begin{equation}
m(T)\approx \pm \pi\sqrt{(T^2_c-T^2)}, \label{mTPQEDTcap2}
\end{equation}
where $T_c^2=\gamma^2\Lambda^2/(2\gamma+\Lambda)^2$. From now on, let us admit that $m(T)>0$. Next, we return $\gamma\rightarrow \alpha T$, to obtain the critical coupling constant
\begin{equation}
\alpha_{c}(T)=\frac{\Lambda}{(\Lambda-2T)}.  \label{ctrT}
\end{equation}

Eq.~(\ref{ctrT}) allows us to estimate an interval for $\alpha_c(T)$. Let us choose $T=0.3 \Lambda$ as an upper bound, because $T/\Lambda \ll 1$. In this case, $\alpha_c(T=0)<\alpha_c<\alpha_c(T=0.3 \Lambda)$, i.e, $1.0<\alpha_c<2.5$. For the particular value $T=0.5\Lambda$ the chiral symmetry is completely restored. However, it is clear that the high-temperature limit $T \approx\Lambda \rightarrow \infty$ is not allowed within this approach. 

Having in mind condensed-matter applications, we consider the fact that the cutoff $\Lambda$ is a physical parameter determined by the lattice spacing, i.e, $\Lambda \propto 1/a_L$, where usually $a_L \approx 10^{-10}$m. For graphene, the cutoff is $\Lambda=\hbar v_F/a_L\approx 3.0$ eV \cite{gra}. Conversely, from a quantum field theory perspective, one should take $\Lambda=\infty$, which implies $\alpha_c=1.0$.

In the case of quantum field theory applications to particle physics, however, dimensional regularization would be more appropriate. In this case, we should return to Eq. (12) and perform the dimensional regularization, generalizing the two-dimensional integral into a D-dimensional integral. Then, after some calculations, we conclude that the critical coupling in the zero-external-momentum approach is equal to one. In fact, this result may also be obtained from 
Eq.~(\ref{eqm}) with $\Lambda=\infty$. Therefore, $\alpha_c(T)=1.0$ is a result that does not depend on the regulator. Nevertheless, the full understanding of the pattern of chiral-symmetry breaking in PQED, using the dimensional regularization, requires a more complete study of the integral equation in Eq.~(\ref{sigmaT}), especially applying better analytical and numerical approaches that go beyond the scope of the present work.

A similar problem has been discussed in the framework of quenched quantum electrodynamics in (3+1)-dimensions (QED4) \cite{Will1,Will2}. For QED4, it has been shown that the results in these two different regulators are in agreement. We shall not discuss more elaborated analytical and numerical approaches by using this regulator in the present work.

We now connect $T_c$ with the mass function at zero temperature, using Eq.~(\ref{mTPQEDTcap2}). For $T=0$, we obtain $T_c\approx m(0)/\pi$. Thus, it is possible to calculate the ratio $R$ between the energy gap $2\,m(0)$ and $T_c$, given by $R=2\pi$. For QED3, the authors in Ref.~\cite{DoreyNL} calculated this ratio and obtained $R_{QED3}=9.36$ for one copy of the Dirac field. 

Eq.~(\ref{mTPQEDTcap2}) shows that the chiral-symmetry breaking in PQED at finite temperatures is a second-order phase transition, with critical exponent equal to $1/2$. 

The zero external-momentum approximation is an ideal approach to include the sum over Matsubara frequencies in Eq.~(\ref{sigmaTn}). Indeed, in this limit we have
\begin{eqnarray}
1=2\gamma \int_0^\Lambda dkk\sum_{n=-\infty}^{+\infty}\frac{1}{(2n+1)^2\pi^2T^2+k^2+m(T)^2} \nonumber \\
\times\frac{1}{\sqrt{k^2+4n^2\pi^2T^2}}. \label{mn}
\end{eqnarray}

We can find numerical solutions to $m(T)$ by using Eq.~(\ref{mn}), thereby verifying that the zero-mode approximation is in good agreement with the SDE for all Matsubara frequencies. This concludes our investigation of chiral-symmetry breaking within the quenched-rainbow approach. The general important points to be retained are: (i) The quenched-rainbow approach provides reasonable physical solutions for the mass function in PQED, similarly to QED4 and quite differently from QED3; (ii) Nontrivial solutions have been found for $T<T_c$; (iii) $T_c$ is estimated from the energy gap at zero temperature; and (iv) The zero-mode approximation is found to be in agreement with the sum over all Matsubara frequencies.

We have assumed that $\Sigma(p,T>T_c)\rightarrow 0$ based on the fact that systems at high temperatures naturally are less prone to quantum effects. Then, no chiral symmetry breaking should be observed. However, for the sake of completeness, we will briefly discuss how to obtain explicitly this trivial solution. From the integral Eq.~(\ref{sigmaT2}), we must use $p^2+T^2+\Sigma^2(p)\approx T^2$, a realistic assumption for high temperatures. Then, by using the expansion in Eq.~(\ref{kernelPQED3}), we find the differential equation
\begin{equation}
\frac{d}{dp}\left(p^2\frac{d\Sigma(p)}{dp}\right)=0, \label{eqdifTgrande}
\end{equation}
the general solution of which is
\begin{equation}
\Sigma(p)=\frac{C_3}{p}+C_4.
\end{equation}

Note that $C_4=0$, since $\Sigma(p\rightarrow \infty)=0$, and $C_3=0$ because $\Sigma(p\rightarrow 0)$ must be finite. In this limit, $T$ is no longer an infrared cutoff, and $\Sigma(p,T)=0$.

\section{Static interactions between massless Dirac particles at $T\neq 0$ in the PQED approach}

In this section, we recalculate the static interaction between charges in the plane,
in order to clarify the physical interpretation of the electromagnetic interaction in the plane at finite temperatures. For this purpose, we must include the corrections of the vacuum polarization to the gauge-field propagator.

From Eq.~(\ref{propA}), we obtain the static gauge-field propagator, given by
\begin{equation}
\Delta_{\mu\nu}(p_0=0,\textbf{p})=\frac{\delta_{\mu 0}\delta_{\nu 0}}{2\sqrt{\textbf{p}^2}}.
\end{equation} 
It has been shown \cite{marino} that this propagator reproduces the Coulomb interactions for static charges in the plane. The static vacuum polarization  at finite temperatures has been calculated in the Ref.~\cite{DoreyNL}, and for massless Dirac particles it is given by
\begin{equation}
\Pi_{\mu\nu}(p_0=0,\textbf{p},T)=-\frac{2\alpha T}{\pi} \delta_{\mu 0}\delta_{\nu 0} f(\textbf{p},T),
\end{equation}
where
\begin{equation}
f(\textbf{p},T)=\int_0^1 dx \ln \left[2\cosh\left(\frac{1}{2}\sqrt{\frac{\textbf{p}^2}{T^2}x(1-x)}\right)\right]. \label{fpt}
\end{equation}
Let us now to consider the high-temperature limit $\textbf{p} \rightarrow 0 $ in Eq.~(\ref{fpt}); then $f(\textbf{p}\rightarrow 0,T)=\ln2$. 

The corrected propagator reads
\begin{equation}
\bar\Delta^{-1}_{\mu\nu}(\textbf{p},T)=\Delta^{-1}_{\mu\nu}(\textbf{p})-\Pi_{\mu\nu}(\textbf{p},T),
\end{equation}
leading to
\begin{equation}
\bar\Delta_{\mu\nu}(\textbf{p},T)=\frac{\delta_{\mu 0}\delta_{\nu 0}}{2\sqrt{\textbf{p}^2}+M^2 (T)},\label{propat2}
\end{equation}
where
\begin{equation}
M^2(T)=\frac{2\alpha T \ln2}{\pi}.
\end{equation}

The static interaction $V(r)$ is obtained from a Fourier transformation of Eq.~(\ref{propat2}),
\begin{equation}
V(r)=e^2\int\frac{d^2 \textbf{p}}{(2\pi)^2} \frac{e^{-i \textbf{p}.\textbf{r}}}{2\sqrt{\textbf{p}^2}+M^2(T)}. \label{defV}
\end{equation}

The angular integral in Eq.~(\ref{defV}) is performed, yielding
\begin{equation}
V(r)=e^2\int_0^{\infty}\frac{dp}{2\pi}\frac{p\,J_0(pr)}{2p+M^2(T)}, \label{defV2}
\end{equation}
where $J_0(pr)$ is the Bessel function and $p=\sqrt{\textbf{p}^2}$. By defining $z \equiv 2p +M^2(T)$ and changing the integration variable into $z$, we obtain, after some algebra, that Eq.~(\ref{defV2}) may be rewritten as
\begin{equation}
V(r)=\frac{e^2}{4\pi r}-e^2\frac{M^2(T)}{4\pi}\int_0^{\infty}dy\frac{J_0(y)}{2y+rM^2(T)}, \label{defV3}
\end{equation}
where $y \equiv p r$. The first term in the rhs of Eq.~(\ref{defV3}) is the Coulomb interaction, while the second one is a correction due to the thermal bath. The integral over $y$ yields \cite{Gradstheyn}
\begin{eqnarray}
\int_0^{\infty}dy\frac{J_0(y)}{2y+rM^2(T)}=\frac{\pi}{4}\left[H_0\left(\frac{r}{r_0}\right)-Y_0\left(\frac{r}{r_0}\right)\right], \label{corcou}
\end{eqnarray}
where 
\begin{equation}
r_0(T)=\frac{2}{M^2(T)}=\frac{\pi}{\alpha \, T \ln2} \label{length}, 
\end{equation}
$H_0(r/r_0)$ is the Struve function, and $Y_0(r/r_0)$ is the Bessel function of the second kind. Therefore, the static interaction between charged particles in PQED in the presence of the thermal bath is
\begin{equation}
V(r,r_0)=\frac{e^2}{4\pi r}\left\{1-\frac{\pi r}{2r_0}\left[H_0\left(\frac{r}{r_0}\right)-Y_0\left(\frac{r}{r_0}\right)\right]\right\}. \label{vstatic}
\end{equation}

We now rewrite the potential $V(r,r_0)$ as a function of the dimensionless variable $l \equiv r/r_0(T)$,
\begin{equation}
r_0V(l)=\frac{e^2}{4\pi l}\left\{1-\frac{\pi }{2}l\left[H_0(l)-Y_0(l)\right]\right\}. \label{vstatic2}
\end{equation}
For $l\ll 1$, we have
\begin{equation}
r_0V(l)= \frac{e^2}{4\pi l}\left[1+l \ln\left(\frac{l}{2\bar\gamma \,}\right)+O(l^2)\right], \label{VPQED}
\end{equation}
where $\ln \bar\gamma=-\gamma_e$ and $\gamma_e$ is the Euler's constant. For $l \gg 1$, we use the asymptotic representation
of the Struve function \cite{Gradstheyn}
\begin{equation}
H_0(l)-Y_0(l)=\frac{1}{\pi} \sum_{m=0}^{p-1}\frac{\Gamma(m+1/2)}{\Gamma(1/2-m)}\left(\frac{l}{2}\right)^{-2m-1}+O(l^{-2p-1}),
\end{equation}
and consider $p=2$, to find 
\begin{equation}
r_0V(l)=\frac{e^2}{4\pi l^3}\left[\pi+O(l^{-2})\right]. \label{vstatic3}
\end{equation}

The last result indicates that the corrected potential, for $r \gg r_0(T)$, goes to zero much faster than the usual Coulomb potential. In Fig.~1, we plot Eq.~(\ref{vstatic}) and the usual Coulomb potential for comparison. 

\begin{figure}[htb]
\label{potecial}
\centering
\includegraphics[scale=0.75]{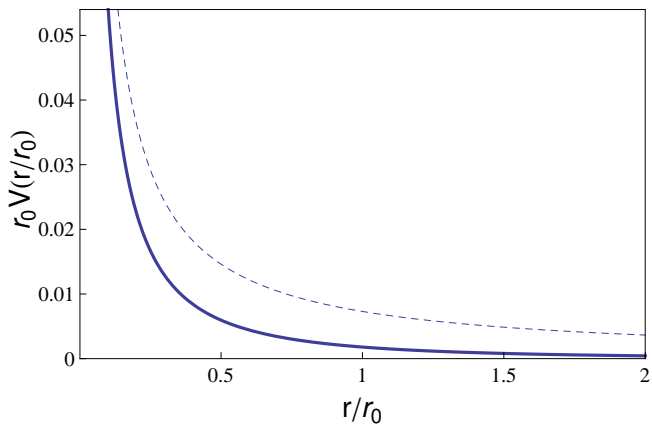}
\caption{(Color online) We artificially consider $\alpha=1/137$ and $r_0=1.0$ for both lines. The thick line is the corrected potential given in Eq.~(\ref{vstatic}), while the dashed line is the usual Coulomb potential for $T=0$. Here $r$ and $r_0$ are given in units of $T^{-1}$.}
\end{figure}

Next, we estimate the physical value of $r_0(T)$. Let us assume a massless particle with energy dispersion $E=v p=\hbar v/r$, where $v= f c$ is the particle velocity, $0\leq f \leq 1$, and $c$ is the light velocity. For graphene $f=1/300$ and for a relativistic particle $f=1$, for instance. Taking this into account, we rewrite Eq.~(\ref{length}) in physical units,
\begin{equation}
r_0(T)=\frac{\pi \hbar f c}{\alpha_f \, k_B T \ln 2},\label{length2}
\end{equation}
where we reintroduced $\hbar$ and $k_B$, the Planck and Boltzmann's constant, respectively and $\alpha_f=\alpha_{QED4}/f=1/(137f)$ is the coupling constant. Thus, $r_0(T)\approx 10^{-5}/T$ m, with $T$ in Kelvin and $f=1/300$. At room temperature $T=300$ K, we find $r_0 \approx 10$ nm, which is much larger than the atomic scale. This value is smaller when $f$ is less than $1/300$. 

There is a parallel between the static interaction in PQED at finite temperatures and QED3 at zero temperature. Indeed, starting from the screened static gauge-field propagator in QED3, obtained from the respective SDE, it has been shown that at large $r$ the static electron-electron interaction is given by $V(r)\propto \ln(e^2 r)+h(r)$, where $h(r) \propto 1/r + O(1/r^2)$ \cite{Roberts2}. Therefore, the results are similar to PQED. However, for QED3 the Coulomb potential is the correction to the confining potential and the term which breaks scale invariance is the dimensional electric charge, instead of the temperature as in PQED. For finite temperatures, the static potential of QED3 behaves like $V(r) \propto 1/\sqrt{ M(T) r} \exp(-M(T) r)$ \cite{DoreyNL}, which is quite different from the corresponding result for PQED in Eq.~(\ref{VPQED}). 

\section{Discussion}

In this paper, we show that the inclusion of finite-temperature effects into the SDE leads to a critical parameter $T_c$ for the onset of dynamical mass generation. Moreover, at high temperatures the chiral symmetry is restored. The finite temperature also increases the value of the critical coupling constant; for low temperatures, we estimate that $1.0<\alpha_c(T)<2.5$, which are large values compared to the equivalent zero-temperature result $\alpha_c(T=0)=\pi/8\approx 0.40$. Our analytical findings were verified by performing numerical tests. We estimate the ratio between the energy gap and the critical temperature for PQED to be $R=2\pi$. Furthermore, we show that the Coulomb potential between static charges in the plane is corrected by a logarithmic potential proportional to $M^2(T)\propto \alpha T$, where $ \alpha=e^2/(4\pi)$ for the short-range limit, while a third-power potential emerges in the long-range limit. A deeper investigation of the ground state generated by this interaction may bring interesting physical results. 

For graphene, the investigation of chiral symmetry breaking in (2+1) dimensional theories is an important topic, since this effect has been related to a dynamical gap generation, which can be relevant to describe electronic transport in this system. It has been argued that this gap energy leads to important technological applications for graphene, such as a graphene transistors \cite{Frank}. PQED at $T=0$ admits gap generation, but only above a critical coupling constant.  Although our results include finite temperatures, and thus are more realistic because any measurement of the gap is done at $T\neq 0$, we are considering $v_F=c$. The inclusion of still more physical parameters to study chiral symmetry breaking in PQED may lead to results closer to the experimental findings in graphene. It may as well bring a deeper understanding of how to control this energy gap. These parameters, among others, should be: the Fermi velocity $v_F$, finite temperature $T$, magnetic field $B$, and chemical potential $\mu$. A recent experimental measurement at low magnetic fields has determined an upper limit of 0.1 meV for a possible gap in suspended monolayer graphene \cite{Elias}. 

According to the Coleman theorem, there is no Goldstone boson in (1+1) dimensions, i.e, no spontaneous (continuos) symmetry breaking may occur \cite{Coleman, Mermin,Hohenberg}. Because of this theorem, one may conclude that there is no ferromagnetism or antiferromagnetism in the Heisenberg model in one or two dimensions \cite{Mermin}. For (2+1) dimensional theories at finite temperatures, the theorem also applies due to the loop integrals, which behave effectively in (1+1) dimensions \cite{DoreyNL}. This is observed for the case of PQED in Eq.~(\ref{sigmaT}). Therefore, there is no \textit{continuous} chiral-symmetry breaking from the very beginning and the dynamically generated mass instead breaks the \textit{discrete} chiral symmetry. For the case of the continuous chiral symmetry, the dynamical mass generation could be related to the Kosterlitz-Thouless mechanism \cite{Kosterlitz}, as has been discussed for the Gross-Neveu model in (2+1) dimensions at finite temperatures \cite{Park}. Because of the Kosterlitz-Thouless mechanism, no continuous symmetry is broken, even in the massive phase, thus avoiding any possible disagreement with the Coleman theorem \cite{Park}.

The non-local approach of PQED has been applied to the study of superconductivity in (2+1) dimensions in the view point of the Kosterlitz-Thouless mechanism \cite{Kovner}. In this case, because of the coupling constant between the matter and the bosonic excitation (lattice effect), it has been shown that the gauge-field propagator has finite mass, thus finite penetration depth, which may lead to an effective description of the Meissner effect in superconductors. Furthermore, we have obtained the ratio $R$ between the energy gap and the critical temperature for the matter field, which is almost twice the value obtained from the BCS theory for superconductivity. We shall explore elsewhere the possible relation between dynamical mass generation, Kosterlitz-Thouless mechanism, and superconductivity \cite{Kovner}.

\section{\textbf{acknowledgments}}

This work was supported in part by CNPq (Brazil), CAPES (Brazil), FAPERJ (Brazil), Programa de Cooperaci\'on Internacional DI10-4002 of the Direcci\'on de Investigaci\'on y Desarrollo de la Universidad de La Frontera (Temuco-Chile) and by the Brazilian
government project Science Without Borders. L. O. Nascimento and V. S. Alves are grateful to M.O.C. Gomes and J. H. A Neto for interesting discussions, L. O. Nascimento is grateful to H.Q. Zhang for critical reading.

\section{Appendix A: Numerical results for the mass function}

\begin{figure}[h]
\begin{center}
\includegraphics[height=5.3cm]{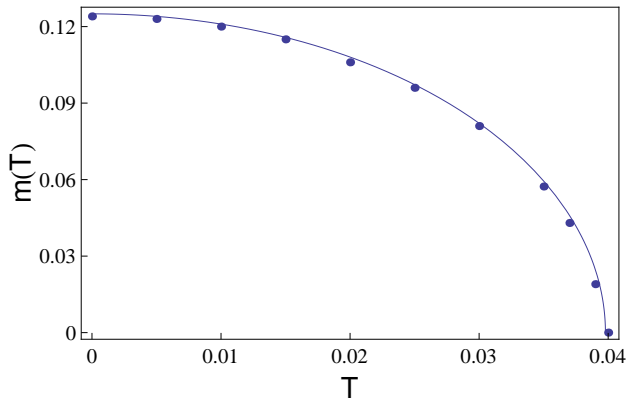}
\caption{\label{massap01} The solid line is the analytical solution given by Eq.~(\ref{mTPQEDTcap2}) and the dots are the numerical results. We use $\Lambda=10$ and $\gamma=1/(8\pi)$ u.$\Lambda$. The critical temperature is $T_c=\gamma=0.04$ u.$\Lambda$.}
\end{center}
\end{figure}

In this appendix, we perform some numerical tests to verify the validity of the analytical approaches adopted in this paper. For simplicity, we will begin by studying Eq.~(\ref{mTn0}) and its analytical solution, Eq.~(\ref{mTPQEDTcap2}), in which for $\Lambda$ large, we obtained $T_c=\gamma$. In Fig.~\ref{massap01}, we plot the numerical results for $\Lambda=10$ and $\gamma=1/(8\pi)$ with dots, and the analytical result as a solid line. Numerical solutions are found by looking for roots of the transcendental equation (\ref{mTn0}). A very good agreement is observed. Note that both solutions vanish almost at the same point $T_c= 0.04$ in units of $\Lambda/10$ (u.$\Lambda$).

Next, we investigated the influence of all Matsubara frequencies for the mass function. After calculating the integral over the momentum  $k$ in Eq.~(\ref{mn}), we find
\begin{eqnarray}
F(m)=1-2\gamma\sum_{n=-\infty}^{+\infty}\frac{1}{\sqrt{m^2+(1+4n)\pi^2T^2}}\times\nonumber \\
\Big[\tan^{-1}\left(\frac{\sqrt{\Lambda^2+4n^2\pi^2T^2}}{\sqrt{m^2+(1+4n)\pi^2T^2}}\right)-\nonumber\\
\tan^{-1}\left(\frac{2n\pi T}{\sqrt{m^2+(1+4n)\pi^2T^2}}\right)\Big]. \label{mTtodosn}
\end{eqnarray}

\begin{figure}[h]
\begin{center}
\includegraphics[height=5.3cm]{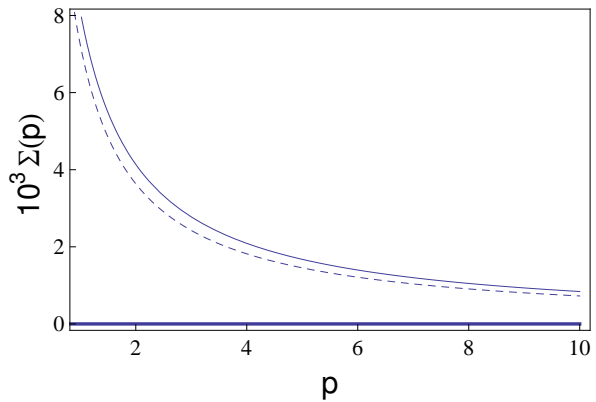}
\caption{\label{massap0} The dashed line is the numerical solution of the nonlinear integral equation given by Eq.(\ref{sigmaT2}). The solid line is the analytical solution Eq.~(\ref{soleqdif}) for $A=0.15$ and $B=0$. For these two lines we used $\Lambda=10$ and $T=0.01$ u.$\Lambda$. The thick line is the numerical solution for $T=0.03$ u.$\Lambda$. For this approach $T_c=0.44\gamma=0.018$ u.$\Lambda$ for $\gamma=1/(8 \pi)$ u.$\Lambda$.} 
\end{center}
\end{figure}

The roots of $F(m(T))=0$ yield the mass function $m(T)$ for a given value of $T$. We estimate the critical temperature by imposing that for $T=T_c$ the mass vanishes, $m(T=T_c)=0$. Thus, we obtain a critical temperature $T_c=0.10$ u.$\Lambda$. This result shows that from the very simple zero-mode approach is already enough to provide an estimate of $T_c$ in units of $\Lambda$. However, a more accurate result is expected when one includes all the Matsubara frequencies. To investigate this more complicated case, we include an effective limit to the Matsubara frequencies, $n\in [-l_o,+l_0]$ with $l_0=10$.  It is possible to show that for any number large than $l_0$, the function $F(m(T))$ remains nearly the same. Indeed, as we demonstrated, the sum over $n$ is convergent, therefore we expect that $u_n=0$ if $n\rightarrow \infty$. 

For non-zero external momentum, we can compare the analytical solution in Eq.~(\ref{soleqdif}) to the numerical solutions of the nonlinear integral Eq.~(\ref{sigmaT2}). For this purpose, we have to choose a value for the arbitrary constant $A$ in the analytical solution. For temperatures larger than $T_c=0.44\gamma=0.018$ u.$\Lambda$ for $\gamma=1/(8\pi)$ u.$\Lambda$, the mass function given by the nonlinear integral equation vanishes, as expected from the analytical solution. Therefore, Fig.~\ref{massap0} shows that the analytical solution found earlier is in very good agreement with the nonlinear integral equation.

\section{Appendix B: The renormalization function $A(p)$}

In this appendix, we show that $A(T,\textbf{p})=1$ is also a suitable approach for finite temperatures. This approach has been discussed for QED3 at zero temperature in Ref.~\cite{Maris2}, while for PQED it has been verified in Ref.~\cite{VLWJF}. After introducing the Matsubara frequencies in Eq.~(\ref{waveAT}) as in Sec. III, we obtain
\begin{eqnarray}
A(\tilde{p})=1&+&\frac{4 \pi \alpha T}{ \tilde{p}^2}\sum^{+\infty}_{n=-\infty}\int\frac{d^2\textbf{k}}{(2\pi)^2}\frac{A(\tilde{k})\Delta(\tilde{q})}{\tilde{k}^2A^2(\tilde{k})+\Sigma^2(\tilde{k})} \nonumber
\\
&\times& \frac{(\tilde{p}.\tilde{q})(\tilde{k}.\tilde{q})}{\tilde{q}^2},
\end{eqnarray}
where $\tilde{p}=(\pi T,\textbf{p})$, $\tilde{k}=((2n+1)\pi T,\textbf{k})$, and $\tilde{q}=(-2n\pi T,\textbf{q})$. We assume $m=0$ because $A(\tilde{p})=A(T,\textbf{p})$ should be independent on Matsubara frequencies. For simplicity, let us consider $\sqrt{\textbf{p}^2}\equiv p$ and $\textbf{p}.\textbf{k}=pk\cos\theta$. The angular integral may be solved analytically, and after some simplifications we obtain
\begin{eqnarray}
I^{\theta}_n(p,k)=\int_0^{2\pi}d\theta \frac{\Delta(\tilde{q})(\tilde{p}.\tilde{q})(\tilde{k}.\tilde{q})}{\tilde{q}^2}=\nonumber \\
\frac{8 n \pi^2 T^2 H_n(p,k) Y_n(p,-k)}{(pk)^{3/2}} E(-4 Y_n(p,k)),
\end{eqnarray}
where $E(x)=\pi/2-(\pi/8) x+...$ for $|x|<1$ is the complete elliptic integral,
\begin{equation}
Y_n(p,k)=\frac{pk}{(p-k)^2+(2n\pi T)^2},
\end{equation} 
and
\begin{eqnarray}
H_n(p,k)=2n Y_n(p,k)^{1/2}(p^2-\pi^2T^2)\nonumber \\
+2pY_n^{1/2}(p,k)(p-k)-pk Y_n^{-1/2}(p,k).
\end{eqnarray}

Next, we obtain the integral equation for $A(\tilde{p})$ with a kernel only dependent on the internal-momentum integral,
\begin{eqnarray}
A(\tilde{p})=1&+&\frac{\alpha T}{ \pi \tilde{p}^2}\sum^{+\infty}_{n=-\infty}\int \frac{k dk A(\tilde{k}) I^{\theta}_n(p,k)}{\tilde{k}^2A^2(\tilde{k})+\Sigma^2(\tilde{k})}. \label{APQEDT}
\end{eqnarray}
As expected, $A(T,\textbf{p})=1+O(\alpha)$, which indicates the accuracy of the approximation $A(p) \approx 1$ \cite{Maris2}. Beyond this argument, we need to look for a numerical solution of Eq.~(\ref{APQEDT}) for $\Sigma(k,T)=0$ (the symmetric phase), as in Fig.~3. Note that, since $\Sigma(k,T)\propto \alpha$, we should not consider $\Sigma(k,T) \neq 0$ in order to solve the integral equation for $A(T,\textbf{p})$ because these new terms are of order  $\alpha^2$. This symmetric phase for $A(T=0,\textbf{p})$ has been also discussed in the context of QED3 \cite{Maris2}.

\begin{figure}[htb]
\label{wave2}
\centering
\includegraphics[scale=0.8]{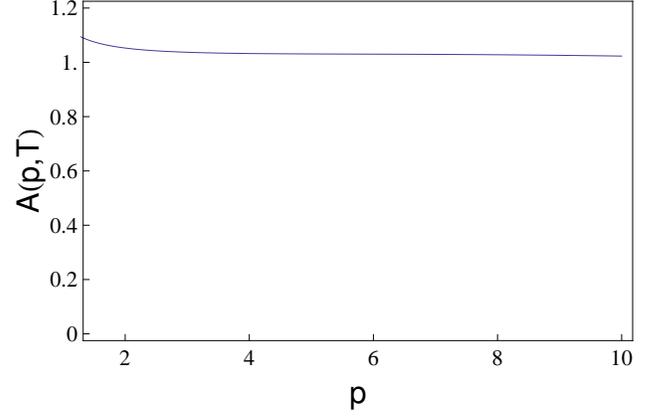}
\caption{(Color online) Numerical solution of Eq.~(\ref{APQEDT}) for $\alpha=1/(8 \pi)$, $T=1.0$ u.$\Lambda$, $\Lambda=10$, $n \in [-10,+10]$, which shows that $A(\tilde{p})\approx 1$ is indeed a realistic approximation.}
\end{figure}

In order to obtain the results in Fig.~1 and Fig.~3, we converted the momentum-dependent kernel in  Eq.~(\ref{sigmaT2}) and Eq.~(\ref{APQEDT}) into a system of nonlinear algebraic equations by using the repeated trapezoidal quadrature rule, as has been successfully done for PQED at $T=0$ \cite{VLWJF}. The fundamental step is to replace the continuum variables $p$ and $k$ into a set of discrete variables $x_i$ and $y_i$, with $10^{-3}<x_i,y_i<\Lambda=10$. In this interval, we obtained a set of $M=300$ solutions for the mass function $\Sigma_i(x_i)$, which implies in a separation $h=(10-10^{-3})/(300-1)\approx 0.033$ between the discrete variables. For $T \neq 0$, the temperature is only a new parameter inside the kernel of the nonlinear integral equations. More details about the numerical procedures may be found in Ref.~\cite{VLWJF} for PQED at $T=0$, in Ref.~\cite{Maris2} for QED3 at $T=0$, and in  Ref.~\cite{Ayala} for QED4 at $T\neq 0$.


\begin{thebibliography}{99}

\bibitem{Roberts} 
C. D. Roberts and A. G. Williams, Nucl. Phys.  \textbf{33}, (1994).

\bibitem{Maris3} 
P. Maris, Phys. Rev. D \textbf{50}, 4189-4193 (1994).

\bibitem{Appel} 
T. Appelquist, M. J. Bowick, E. Cohler, and L. C. R. Wijewardhana, Phys. Rev. Lett. \textbf{55}, 1715 (1985).

\bibitem{Kondo} 
K. I. Kondo, P. Maris, Phys. Rev. D \textbf{52}, 1212-1228 (1995).

\bibitem{Maris2} 
P. Maris, Phys. Rev. D \textbf{54}, 4049-4057 (1996).

\bibitem{Hong} 
D. K. Hong and S. H. Park, Phys. Rev. D \textbf{47}, 3651-3654 (1993).

\bibitem{Ayala} 
A. Ayala, A. Bashir, Phys. Rev. D \textbf{67},  076005 (2003).

\bibitem{Dorey} 
N. Dorey and N. E. Mavromatos, Phys. Lett. B \textbf{266},163-168 (1991).

\bibitem{Aitchison} 
I. J. R. Aitchison and M. Klein-Kreisler, Phys. Rev. D \textbf{50}, 1068 (1994).

\bibitem{Triant} 
G. Triantaphyllou, Phys. Rev. D \textbf{58}, 065006 (1998).

\bibitem{Oskar} O. Vafek and A. Vishwanath, Ann. Rev. Cond. Matt. Phys. {\bf 5}, 83-112 (2014).

\bibitem{GeimIQHE} K. S. Novoselov, A. K. Geim, S. V. Morozov, D. Jiang, M. I. Katsnelson, I. V. Grigorieva, S. V. Dubonos, and A. A. Firsov, Nature (London) {\bf 438}, 197 (2005).

\bibitem{MariaVoz} M. A. H. Vozmediano and F. Guinea. Phys. Scr. {\bf T146}, 014015 (6pp) (2012).

\bibitem{Drut} J. E. Drut and T. A. Lahde, Phys. Rev. B {\bf 79}, 165425 (2009).

\bibitem{ChunXu} C. X. Zhang, G. Z. Liu, and M. Q. Huang, Phys. Rev. B {\bf 83}, 115438 (2011).

\bibitem{VLWJF} V. S. Alves, W. S. Elias, L. O. Nascimento, V. Juri\v ci\' c and F. Pe\~na, Phys. Rev. D {\bf 87}, 125002 (2013).

\bibitem{review} A. H. Castro Neto, F. Guinea, N. M. R. Peres, K. S. Novoselov and A. K. Geim, Rev. Mod. Phys. {\bf 81}, 109 (2009).

\bibitem{marino} E. C. Marino, Nucl. Phys. B {\bf 408}, 551 (1993).

\bibitem{DoreyNL} N. Dorey and N. E. Mavromatos, Nucl. Phys. B {\bf 386}, 614 (1992).

\bibitem{Kovner} A. Kovner and B. Rosenstein, Phys. Rev. B {\bf 42}, 4748 (1990).

\bibitem{marino1} E. C. Marino, Phys. Lett. B {\bf 263}, 63 (1991).

\bibitem{Miransky0} E. V. Gorbar, V. P. Gusynin and V. A. Miransky, Phys. Rev. D {\bf 64}, 105028 (2001).

\bibitem{Teber} S. Teber, Phys. Rev. D {\bf 86}, 025005 (2012).

\bibitem{Barci} 
D. G. Barci and L. E. Oxman, Mod. Phys. lett. A \textbf{12}, 493-500 (1997).

\bibitem{Marino2} 
R. L. P. G. do Amaral e E. C. Marino, J. Phys. A: Math. Gen. \textbf{25}, 5183-5200 (1992).

\bibitem{Barcelos} 
R. Amorim and J. Barcelos Neto, J. Math. Phys. \textbf{40}, 585-600 (1999).
\bibitem{huygens} C. G. Bollini and J. J. Giambiagi, J. Math. Phys. (N.Y.) {\bf 34}, 610 (1993).

\bibitem{unitarity} E. C. Marino, L. O. Nascimento, V. S. Alves, and C. M. Smith , Phys. Rev. D {\bf 90}, 105003 (2014).

\bibitem{Abrikosov} 
A. A. Abrikosov, L. P. Gorkov, I. E. Dzyaloshinski. \textit{Methods of Quantum Field Theory in Statiscal Physics}, Dover Publications, New York 1963; Ashok Das. \textit{Finite Temperature Field Theory}.
World Scientific, (1997).

\bibitem{Arfken} 
G. Arfken. \textit{Mathematical Methods for Physicists}. Third Edition,
Academic Press, (1985).

\bibitem{vterm} J-P. Blaizot and U. Reinosa, Nucl. Phys. A {\bf 764}, 393-422, (2006).

\bibitem{Gradstheyn} 
I. S. Gradstheyn and I. M.Ryzhik. \textit{Table of Integrals, Series and Products}. Seventh Edition,
Academic Press, (2007).

\bibitem{gra} E. C. Marino,  L. O. Nascimento, V. S. Alves, and C. M. Smith, Phys. Rev. X {\bf 5 }, 011040, (2015). 

\bibitem{Will1} A. W. Schreiber, T. Sizer, and A. G. Williams, Phys. Rev. D {\bf 58}, 125014, (1998).

\bibitem{Will2} V. P. Gusynin, A. W. Schreiber, T. Sizer, and A. G. Williams. Phys. Rev. D {\bf 60}, 065007, (1999).

\bibitem{Roberts2} C. J. Burden, J. Praschifka, and C. D. Roberts, Phys. Rev. D. {\bf 46}, 2695-2702, (1992).

\bibitem{Frank} F. Schwierz, Nat. Nanotech. {\bf 5}, 487-496, (2010).

\bibitem{Elias} D. C. Elias, R. V. Gorbachev, A. S. Mayorov, S. V. Morozov, A. A. Zhukov, P. Blake, L. A. Ponomarenko, I. V. Grigorieva, K. S. Novoselov, F. Guinea, and A. K. Geim, {\it Nat. Phys. 7}, {\bf 701} (2011).

\bibitem{Coleman} S. Coleman, Commun. Math. Phys. {\bf 31}, 259-264 (1973).

\bibitem{Mermin} N. D. Mermin and H. Wagner, Phys. Rev. Lett. {\bf 17}, 1133-1136 (1966).

\bibitem{Hohenberg} P. C. Hohenberg, Phys. Rev. {\bf 153}, 493 (1967).

\bibitem{Kosterlitz} J. M. Kosterlitz and D. J. Thouless, J. Phys. C {\bf 6}, 1181 (1973).

\bibitem{Park} S. H. Park, B. Rosenstein, and B. J. Warr, SLAC preprint: SLAC-PUB-5349 (1990).

\end{thebibliography}
\end{document}